\documentclass[conference]{IEEEtran}
\IEEEoverridecommandlockouts
\usepackage{cite}
\usepackage{amsmath,amssymb,amsfonts}
\usepackage{algorithmic}
\usepackage{graphicx}
\usepackage{textcomp}
\usepackage{xcolor}
\def\BibTeX{{\rm B\kern-.05em{\sc i\kern-.025em b}\kern-.08em
    T\kern-.1667em\lower.7ex\hbox{E}\kern-.125emX}}

\usepackage{bm}
\usepackage{tikz}

\usepackage[ruled,vlined]{algorithm2e}
\usepackage{float}
\newcommand{\gvec}[1]{\bm{\mathit{#1}}}

\usepackage{tabularx}
\usepackage{booktabs}

\usepackage{multirow}
\usepackage{multicol}

\newcommand\inputpgf[2]{{
		\let\pgfimageWithoutPath\pgfimage
		\renewcommand{\pgfimage}[2][]{\pgfimageWithoutPath[##1]{#1/##2}}
		\input{#1/#2}
}}
\usepackage{import}
\graphicspath{{./images/}}

\begin{document}

\title{Unsupervised Clustered Federated Learning in Complex Multi-source Acoustic Environments\\
\thanks{This work has been supported by the German Research Foundation (DFG) - Project Number 282835863.}
}

\author{\IEEEauthorblockN{Alexandru Nelus, Rene Glitza, and Rainer Martin}
\IEEEauthorblockA{\textit{Institute of Communication Acoustics}, 
\textit{Ruhr-Universit\"at Bochum}\\
44780 Bochum, Germany \\
firstname.lastname@rub.de}
}

\maketitle

\begin{abstract}

In this paper we introduce a realistic and challenging, multi-source and multi-room acoustic environment and an improved algorithm for the estimation of source-dominated microphone clusters in acoustic sensor networks. Our proposed clustering method is based on a single microphone per node and on unsupervised clustered federated learning which employs a light-weight autoencoder model. We present an improved clustering control strategy that takes into account the variability of the acoustic scene and allows the estimation of a dynamic range of clusters using reduced amounts of training data. The proposed approach is optimized using clustering-based measures and validated via a network-wide classification task.



\end{abstract}



\section{Introduction}
\label{sec:introduction}

Acoustic sensor networks (ASNs) are gaining increasing popularity thanks to their capacity to exploit information from various (wireless) and sometimes ad-hoc acoustic sensors. ASNs have become hosts to many applications \cite{bertrand2011applications, pasha2019survey}, e.g., source localization \cite{gannot_kellermann}, event classification \cite{ebbers9022638}, speech enhancement \cite{MARKOVICHGOLAN20154}, etc. These, in turn, can benefit from knowledge of \mbox{source-dominated} clusters formed by microphone-carrying nodes in the vicinity of distributed acoustic sources \cite{Gergen_beamforming_2018, bahari2017distributed}.


To support these emerging applications we propose an improved solution for unsupervised clustering of ad-hoc sensor nodes based on clustered federated learning (CFL) \cite{Sattler9174890} which is derived from federated learning (FL) \cite{McMahanMRA16}. The latter can be considered as privacy-preserving, decentralized collaborative machine learning \cite{yang2019federated}. In this approach, ASN nodes (clients) locally train a neural network (NN) model and solely share locally learned NN weight updates with a central node (server). In this way, it is ensured that audio data is confined to the node level. Moreover, these NN weight updates are aggregated by the server and the new master model is downloaded by the clients. Clusters are then formed on the basis of weight update similarities and the CFL cycle can be further repeated inside each cluster to yield sub-clusters.


In our approach, the (\textit{unsupervised}) light-weight autoencoder at node level provides the flexibility to adapt online to varying and unseen acoustic conditions, room configurations, speakers, noise sources etc. Exploiting this advantage, we propose pre-training of the autoencoder before CFL, freezing all layers except the bottleneck layer, and only updating the latter during CFL. This allows a drastic reduction of the number of NN weight updates sent to and aggregated by the server, along with avoidance of overfitting issues that come along with small training sets.

For the evaluation of the proposed clustering method we introduce a complex acoustic environment, based on the apartment layout described in \cite{Dekkers2017}, with multiple rooms that have different reverberation conditions, open doors, furniture, and decorations along with multiple ($\geq2$) sources of interest and without prior information about their number. In order to handle these challenging conditions, novel CFL control criteria are proposed that take into consideration the high acoustic variability of each cluster and allow the estimation of clusters using only 40 s of audio data. Moreover, the computation of cluster membership values is also updated in order to handle the variety of clusters generated. The evaluation methods consist of distance-based measures and an application in the form of a \mbox{network-wide} gender recognition task.

The remainder of this paper is structured as follows: we present the relation to both state-of-the-art and own prior work, after which we detail the proposed methods. This is followed by a description of the experimental results along with a discussion of results and conclusions.

\section{Relation to prior work}
\label{sec:priorwork}
Microphone clustering in ASNs has been previously explored using e.g., coherence-based features \cite{himawan2010clustering, Pasha2017}, eigenvectors \cite{bahari2017distributed}, divergence of power spectral densities \cite{zhao2020model} or cepstral features \cite{Gergen_journal_2015, Gergen_beamforming_2018}. These clustering solutions and applications, although effective, do not explicitly incorporate privacy considerations and evaluations are confined to shoebox-type room scenarios.


CFL \cite{Sattler2019} has been mainly demonstrated in (semi-) supervised learning applications with (weak) classification labels and relatively large data sets  \cite{Sattler9174890, SattlerWMS19, SattlerMWS20}. Moreover, hard-clustering without the possibility of generating membership values has been used. We have recently addressed some of the aforementioned aspects and successfully explored the application of an unsupervised CFL clustering scheme in ASNs \cite{Nelus2021}. However, the latter work is also limited to a shoebox room with only two simultaneously active sources and, consequently, only two clusters. The number of sources was considered to be known and was used as a stopping criterion for the clustering algorithm. Estimation of clusters was performed on audio segments of 160 s.



\section{Unsupervised clustered federated learning}

Data submitted by distributed clients to a collaborative training system can be non-i.i.d. Therefore, a single global vector of model parameters $\gvec{\theta}^*$ may not be able to minimize the loss of all clients at the same time. This impediment has led to the development of the {\itshape clustered} FL approach \cite{Sattler9174890}. CFL groups clients with \textit{congruent} data distributions into individual clusters such that each cluster $c$ learns its own model parameters $\gvec{\theta}_{c}$. To this end, we define a matrix containing the $L_2$-normalized weight update vectors of all M nodes, $\gvec{\Delta\theta} = \left(\gvec{\Delta\theta}_i / \lVert \gvec{\Delta\theta}_i  \rVert , \ldots, \gvec{\Delta\theta}_M / \lVert \gvec{\Delta\theta}_M \rVert\right)$ and compute the {\itshape cosine similarity matrix} $\gvec{A} = \gvec{\Delta\theta}^T \gvec{\Delta\theta}$ between the nodes' weight update vectors. \mbox{Bi-partitioning} the set of nodes results in clusters $c_1$ and $c_2$, such that the minimum intra-cluster similarity is larger than the maximum inter-cluster similarity. CFL along with the generation of a new cosine similarity matrix $\gvec{A}_c$, are performed for each new cluster $c$. Bi-partitioning is further applied, if needed, and the process is repeated until the data congruence condition is satisfied. Clusters with only two nodes are no longer bi-partitioned.


In order to assess the congruence property within a cluster $c$, we make use of the mean and the maximum Euclidean norms of the clients' weight update vectors, defined as \cite{Sattler9174890}
\vspace{-0.05 cm}
\begin{equation}
    \Delta\bar{\theta}_c = \left\| \frac{1}{|c|} \sum_{i \in c} \gvec{\Delta\theta}_i \right\| \quad\mathrm{and}\quad \Delta\hat{\theta}_c = \max_{i \in c} (\| \gvec{\Delta\theta}_i \|).
    \vspace{-0.05 cm}
\label{eq:mean_max_updates}
\end{equation}
In the standard CFL approach, \mbox{$\Delta\bar{\theta}_c \leq \varepsilon_1$} and $\Delta\hat{\theta}_c \geq \varepsilon_2$, where $\varepsilon_1$ and $\varepsilon_2$ are empirically set, indicate that the system has reached a stationary solution but this is not optimal for all clients (large individual gradients), thus prompting \mbox{bi-partitioning}. To account for varying cluster sizes and absolute values of weight update vectors caused by differing audio sources and room acoustics, we propose to use a dynamic initialization of $\varepsilon_1$ as a weighted sum of $\Delta\bar{\theta}_c$ and $\Delta\hat{\theta}_c$. This is done at cluster level in the CFL communication round $\tau=0$. Additionally, we replace the second condition $\Delta\hat{\theta}_c \geq \varepsilon_2$ by the ratio between the mean and maximum norms $ \Delta\bar{\theta}_c/\Delta\hat{\theta}_c \leq \varepsilon_2$ and thereby arrive at a more robust normalized test criterion.

The aforementioned congruence assessment is performed after a minimum of $\tau > min_{\tau}$ communication rounds and the entire algorithm is stopped after $max_{\tau}$ rounds. Furthermore, an additional stopping criterion $n_\mathrm{no\text{-}split} > \varepsilon_3$ is proposed in order to account for the weight updates' divergence caused by training for too many consecutive communication rounds on congruent data. The CFL algorithm, along with the proposed modifications, is comprehensively detailed in Algorithm \ref{alg:ucfl}. 

\begin{algorithm}[t!]
	\SetAlgoLined
	\KwIn{Pre-trained autoencoder $h$, thresholds $\varepsilon_2$ and $\varepsilon_3$, lower ($\mathit{min}_\tau$) and upper ($max_{\tau}$) communication rounds bounds}

	Freeze all parameters of $h$ except bottleneck subset $\gvec{\theta}$\\
    \While{\text{audio buffer $!=$ empty}}{ 
    Read audio data $\gvec{D}$ of M clients \\
    Initialize cluster list $C \leftarrow \{\{1,..M\}\}$ with a single cluster element that contains all $M$ clients\\
    $\gvec\theta_c \leftarrow \gvec\theta \leftarrow$ \text{random initialization}, $\forall c \in C$\\
	$n_\mathrm{no\text{-}split}\leftarrow -1$, $\tau \leftarrow 0$\\
	\While{$\tau \leq \mathit{max}_\tau \ \text{\normalfont \textbf{and}}\  n_\mathrm{no\text{-}split} \leq \varepsilon_3$}{
		\For {$i \leftarrow 1,\dots,M$ \textbf{\textit{in parallel}}}{
			\underline{Client i does:}\\
$			\gvec{\theta}_i^{\tau} \leftarrow \gvec{\theta}_{c}^{\tau}, \ i \in c \text{ and } c \in C$\\
			$\Delta \gvec{\theta}_i^{\tau} \leftarrow \mathrm{SGD} (\gvec{\theta}_i^{\tau}, \gvec{D}_i) - \gvec{\theta}_i^{\tau}$\\
		}
		\underline{Server does:}\\
		$C' \leftarrow \{ \}$\\
		$\mathit{split} \leftarrow \mathrm{false}$\\
		\For{$c \in C$}{
			Compute $\gvec{A}_c$ and $\Delta\bar{\theta}_c, \Delta\hat{\theta}_c$ from (\ref{eq:mean_max_updates})
			
			$\text{\textbf{if}}\ \tau == 0 \text{\textbf{ then}}\ \varepsilon_1 \leftarrow \Delta\bar{\theta}_c + 0.1 \Delta\hat{\theta}_c$ \\
			\eIf{$\tau > \mathit{min}_\tau \ \text{\normalfont \textbf{and}}\  \Delta\bar{\theta}_c \leq \varepsilon_1 \ \text{\normalfont \textbf{and}}\ \frac{\Delta\bar{\theta}_c}{\Delta\hat{\theta}_c} \leq \varepsilon_2$}{
				$\{c_1, c_2\} \leftarrow \text{Bi-partitioning}(c, \gvec{A}_c)$ \\	
				$\gvec{\theta}^{\tau+1}_{c_x} \leftarrow \gvec{\theta}^{\tau}_c + \frac{1}{|c_x|}\sum_{i \in c_x} \Delta\gvec{\theta}^{\tau}_{i}, \ \forall c_x \in \{c_1, c_2\}$\\
				$\mathit{split} \leftarrow \mathrm{true}$\\
				$n_\mathrm{no\text{-}split} \leftarrow 0$\\
				$C' \leftarrow C' + \{c_1, c_2\}$\\
			}
			{
				$\gvec{\theta}^{\tau+1}_{c} \leftarrow\gvec{\theta}^{\tau}_c + \frac{1}{|c|}\sum_{i \in c} \Delta\gvec{\theta}^{\tau}_{i}$\\
				$C' \leftarrow C' + \{c\}$\\
			}
		}
		\If{{\normalfont\textbf{not}} $\mathit{split}$ {\normalfont \textbf{and}} $n_\mathrm{no\text{-}split} \geq 0$}{
			$n_\mathrm{no\text{-}split} \leftarrow n_\mathrm{no\text{-}split}+1$\\
		}
		$\tau \leftarrow \tau+1$ \\
		$C \leftarrow C'$ \\
	}
	}
	\caption{Unsupervised CFL for the estimation of source-dominated microphone clusters in ASNs.}
	\label{alg:ucfl}
\end{algorithm}

\subsection{Autoencoder description}

One of the necessary elements in adapting CFL to ASN scenarios that lack training labels is the incorporation of an unsupervised model, namely an autoencoder, thus leading to unsupervised CFL. Moreover, the short time intervals on the basis of which a clustering estimation is desired greatly limit the amount of training data clients can train on, potentially leading to overfitting issues. For this reason, \cite{Nelus2021} proposes pre-training an autoencoder $h$ on more extensive data and then freezing all its layers except the bottleneck layer. The latter can be re-initialized with random values and further adapted with much smaller amounts of data by individual clients. 

Table \ref{tab:autoencoder} summarizes the model, with the fifth layer being the bottleneck layer. To train the model, we use the mean squared error (MSE) loss between the input $\gvec{Y}$ and reconstructed $\gvec{\hat{Y}}$ log-mel band energy (LMBE) feature vectors.


\begin{table}[t]
\centering
\caption{Neural network architecture of autoencoder $h$.}
\vspace{0.15cm}
\label{tab:autoencoder}
\resizebox{\columnwidth}{!}{%
\begin{tabular}{c|c|c|c|c|c|c}
\hline
Layer & Input         & Operator    & \begin{tabular}[c]{@{}c@{}}Out\\ ch.\end{tabular} & Stride & \begin{tabular}[c]{@{}c@{}}Kernel/\\ Nodes\end{tabular} & Activation \\ \hline
1     & 128 x 128     & Conv2d      & 6                                                 & 1      & 5 x 5                                                   & ReLu       \\
2     & 6 x 124 x124  & MaxPool     & -                                                 & 2      & 2 x 2                                                   & -          \\
3     & 6 x 62 x 62   & Conv2d      & 16                                                & 1      & 5 x 5                                                   & ReLu       \\
4     & 16 x 58 x 58  & MaxPool     & -                                                 & 2      & 2 x 2                                                   & -          \\ \hline
5     & 16 x 29 x 29  & Dense       & -                                                 & -      & 29                                                      & ReLu       \\ \hline
6     & 16 x 29 x 29  & Unpool      & -                                                 & 2      & 2 x 2                                                   & -          \\
7     & 16 x 58 x 58  & ConvTrans2d & 6                                                 & 1      & 5 x 5                                                   & ReLu       \\
8     & 6 x 62 x 62   & Unpool      & -                                                 & 2      & 2 x 2                                                   & -          \\
9     & 6 x 124 x 124 & ConvTrans2d & 1                                                 & 1      & 5 x 5                                                   & Sigmoid   
\end{tabular}%
}
\vspace{-0.05 cm}
\end{table}
\subsection{Cluster membership values}
\label{sec:mem_values}

As observed in fuzzy clustering approaches \cite{Gergen_journal_2015}, determining cluster membership values (MVs) offers vital benefits to ASN-based applications. While standard CFL provides only hard clustering results, we have previously extended CFL to also compute MVs  \cite{Nelus2021}. We now refine this method in order to cope with more than two sources and two clusters. 




For each cluster $c \in C$, we first compute the mean intra- and inter-cluster similarities of each client $i \in c$ using $\gvec{A}_c$ and stack them in vectors $\gvec{q}$ and $\gvec{r}$, respectively, with
\vspace{-0.1 cm}
\begin{equation}
    q_i = \frac{1}{|c|-1} \sum_{j \in c \setminus \{i\}} A_{i,j} \enskip\mathrm{and}\enskip r_i = \frac{1}{M-|c|}  \sum_{k \in C^* \setminus c} A_{i,k},
\vspace{-0.1 cm}
\end{equation}
where $|\cdot|$ denotes the cardinality of a set, $C$ is the total set of clusters, $M$ is the total number of clients and $C^* = \{1,..,M\}$.
Next, we apply min-max normalization to the intra- and inter-cluster similarities $\gvec{q}$ and $\gvec{r}$, and balance them in vector $\gvec{p}$,
\vspace{-0.3 cm}
\begin{equation}
\label{eq:lambda}
    \gvec{p}=\lambda \gvec{q} + (1 - \lambda) \gvec{r},
\vspace{-0.15 cm}
\end{equation}
with elements $p_i$. This is motivated by the observation that nodes closest to a source have not only a small inter-cluster similarity but also a small intra-cluster similarity. 
The node with the smallest $p_i$ value is then selected as the cluster's reference node, assuming this is the most representative node for the source. The min-max normalized cosine similarities between the latter and all other nodes form the cluster's MVs vector $\gvec{\mu}$. Thresholding with $\mu_i=0, \forall \mu_i \leq v$ is applied in order to disregard nodes with low MVs. The impact of $\lambda$ and the MV thresholding are further studied in Section \ref{sec:experiments}.




\section{Experimental setup and results}
\label{sec:experiments}

\begin{figure}[t]
    \centering

    \resizebox{0.75\columnwidth}{!}{%
		\def\svgwidth{1.0\textwidth}
		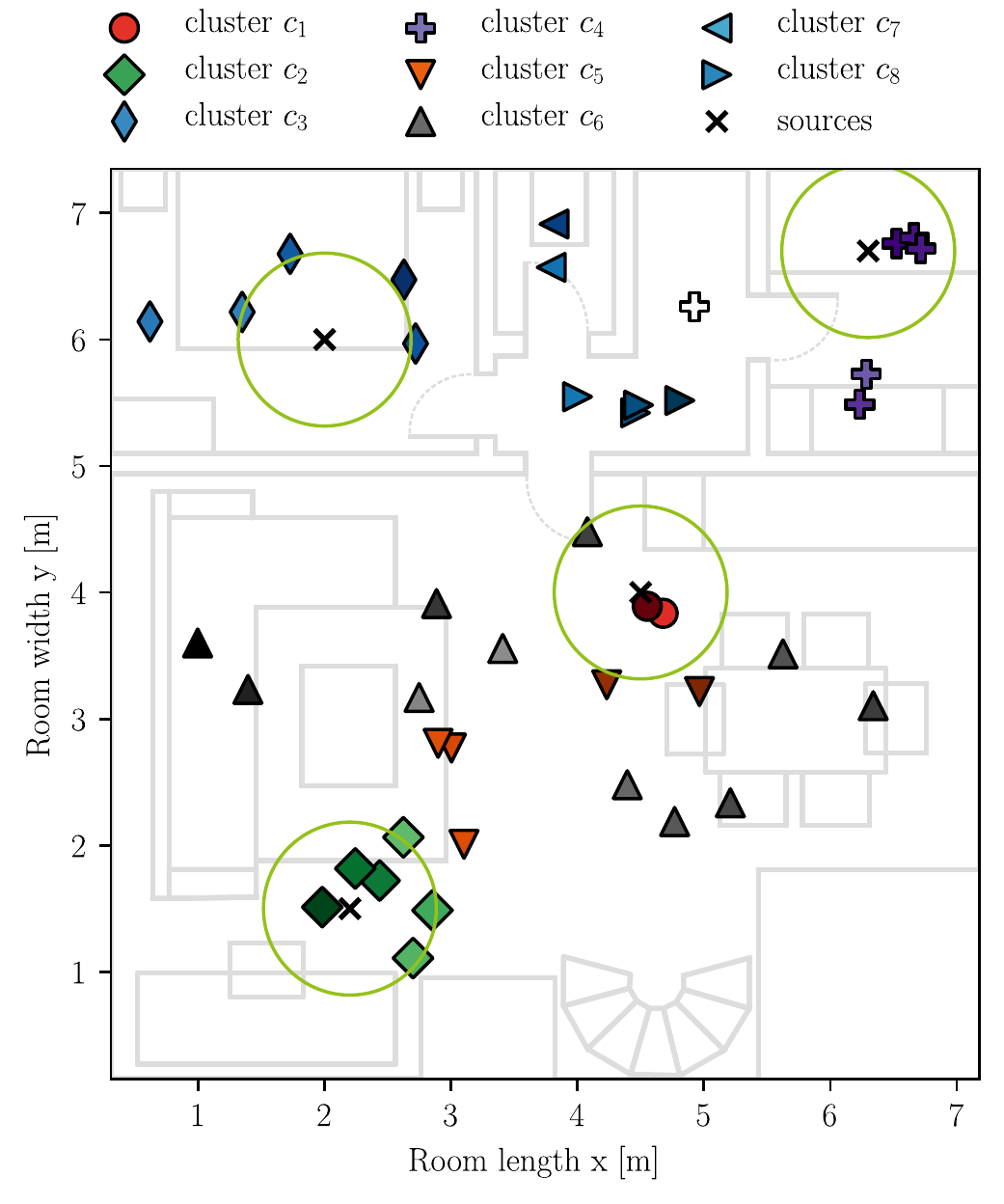
	}
	\vspace{-0.3cm}
    \caption{Floor plan of simulated SINS apartment with cluster estimations for a single scenario. Color intensity is proportional to cluster membership values and green circles indicate the aggregated critical distance.}
    \label{fig:apartment_layout}
\end{figure}

\subsection{Database and rooms}
\label{sec:database}
\begin{table}[b]
    \vspace{-0.1 cm}
	\centering
	\caption{Room-specific floor area, simulated reverberation time ($T_{60}$) and number of randomly positioned sources and nodes.}
	\resizebox{\columnwidth}{!}{%
	\begin{tabular}{l|cccc}
		\toprule
		Room        & Floor Area   & $T_{60}$	& \# Sources & \# Nodes	  \\ \midrule
		Living room & 29.14 m$^2$  & 0.37 s	    & 2           & 24    \\
		Bedroom     & 6.51 m$^2$   & 0.27 s	    & 1           & 5    \\
		Toilet      & 1.22 m$^2$   & 0.33 s	    & 0           & 2    \\
		Corridor    & 2.86 m$^2$   & 0.42 s	    & 0           & 5    \\
		Bathroom    & 3.77 m$^2$   & 0.53 s	    & 1           & 5    \\ 
	\end{tabular}%
	}
	\label{tab:sins_t60}
	\vspace{-0.2 cm}
\end{table}

We employ a subset of the LibriSpeech corpus \cite{panayotov2015librispeech}, namely \textit{train-clean-100}, with 251 speakers (125 female, 126 male) extracted from 16 kHz audiobook recordings. Voice activity detection (VAD) is applied and the data is restructured into 25006 utterances of length 10 s each. The data is further split into \textit{Libri-server} with 157 speakers (79 female, 78 male) for autoencoder and gender recognizer training, and \textit{Libri-clients} (94 speakers) for clustering and speaker gender inference.

The apartment layout introduced in \cite{Dekkers2017} and schematically illustrated in Fig. \ref{fig:apartment_layout}, is transformed into a 3D model. Auralization using CATT-Acoustic with cone-tracing \cite{DalenbaeckTUCT2019} is then performed. Open doors, typical furniture, utensils, and decorations are included. The respective $T_{60}$ reverberation times along with the rooms' floor areas, number of randomly positioned nodes, and simultaneously active sources are presented in Table \ref{tab:sins_t60}. The random positioning of nodes is made under the constraint that for every source in the living room, a minimum of three nodes are within critical distance, thus having higher direct component energy than reverberation energy. The audio signal $x_i(t)$ captured by each ASN node $i$ is expressed as
\vspace{-0.5cm}
\begin{equation}
\label{eq:IRs}
    x_i(t) = \sum_{z=0}^{N_S} s_z(t)*g_{z,i}(t),
\vspace{-0.05cm}
\end{equation}
where $g_{z,i}$ is the room impulse response (RIR) from source $s_z$ to node $i$ and $N_S$ is the scenario's total number of sources.

In order to analyze the generalization of our proposed clustering method, we consider two types of ASN scenes. The first, \textit{2SL}, only involves the living room which contains two simultaneously active sources (male and female). The second, \textit{4SA}, involves the entire apartment and contains four simultaneously active sources (two male and two female). For each ASN scene, ten randomly positioned source-node constellations are created and for each constellation 20 gender-balanced speaker pairs are randomly selected from Libri-clients resulting in 200 simulation scenarios. We randomly select four utterances/speaker (10 s each) to perform CFL followed by gender recognition using the estimated clusters.
\vspace{-0.1 cm}

\subsection{Autoencoder pre-training}
\label{sec:autoenc_train}

The autoencoder $h$ is pre-trained on the Libri-server set for 300 epochs using an SGD optimizer with a learning rate of $l_r=0.1$. For each 10 s utterance, an LMBE feature vector $\gvec{Y}$ is extracted, as detailed in \cite{Nelus2019variational}, using a short-time discrete Fourier transform (STFT) with window length $L_1=0.064$ s and step size $R_1=0.032$ s, along with $K=128$ mel filters.

\subsection{Clustering}
\label{sec:clustering_exp}

\begin{table}
    \caption{Normalized cluster-to-source distance $\Tilde{d}_{c_x}^{s_z}$ from cluster $c_x$  to source $s_z$, averaged over 200 scenarios. Results include entire SINS apartment (4SA) and living room only (2SL).}
	\centering
	\resizebox{\columnwidth}{!}{%
	\begin{tabular}{l|ccccccccccc}
		\toprule
		  4SA    &     $c_1$     &     $c_2$     &     $c_3$     &     $c_4$     & $c_5$ & $c_6$ & $c_7$ & $c_8$ & $c_9$ & $c_{10}$ & $c_{11}$ \\ \midrule
		$s_1$ & \textbf{0.20} &     0.84      &     0.81      &     0.97      & 0.44  & 0.71  & 0.77  & 0.72  & 0.72  & 0.53   & 0.69   \\
		$s_2$ &     0.81      & \textbf{0.24} &     1.19      &     1.11      & 0.92  & 0.69  & 1.12  & 1.12  & 1.07  & 0.92   & 1.65   \\
		$s_3$ &     0.81      &     1.04      & \textbf{0.11} &     1.01      & 0.68  & 0.75  & 0.46  & 0.46  & 0.77  & 0.82   & 0.79   \\
		$s_4$ &     0.98      &     1.01      &     1.00      & \textbf{0.08} & 0.83  & 0.80  & 0.63  & 0.63  & 0.42  & 0.82   & 1.01   \\ \midrule
		
			2SL  &     $c_1$     &     $c_2$     & $c_3$ & $c_4$ & $c_5$ & $c_6$\\ \midrule
		$s_1$ & \textbf{0.08} &     0.95      & 0.37  & 0.78  & 0.48 & 0.76 \\
		$s_2$ &     0.94      & \textbf{0.09} & 0.68  & 0.28  & 0.67 & 0.30 \\
	\end{tabular}%
	}
	\label{tab:cts_grouped}
\end{table}

The autoencoder $h$ described above is used for unsupervised CFL as indicated in Algorithm \ref{alg:ucfl}. All the $\gvec{\Theta}$ parameters, except for the subset $\gvec{\theta}$, that corresponds to the bottleneck layer, are frozen. The latter is randomly re-initialized before each clustering estimation which is now performed for four utterances (total of 40 s). This reduces the initial number of trainable parameters from $O_1 = 5999$ to $O_2 = 841$. The CFL control criteria are set to $\varepsilon_2=0.84$ and $\varepsilon_3=2$.

In order to evaluate clustering performance in relation to previous work \cite{Nelus2021} and state-of-the-art implementations \cite{Gergen_journal_2015}, we extend the normalized cluster-to-source distance (CTS) $\tilde{d}^{s_z}_{c_x}$ of cluster $c_x$ to source $s_z$ \cite{Gergen_journal_2015} to multiple sources and clusters,
\begin{equation}
\vspace{-0.05 cm}
	\tilde{d}^{s_z}_{c_x} = \frac{\| \rho_{s_z} - \overline{\rho}_{c_x} \|}{\overline{d}_{S}}, 
	\ \forall c_x \in C \ \text{and}\ s_z \in S,
	\label{eq:cts}
\end{equation}
where $\rho_{s_z}$ is the position of source $s_z$. The centroid $\bar{\rho}_{c_x}$ represents the average of geometric positions of nodes $i$ assigned to cluster $c_x$ weighted by their respective MVs, $\overline{d}_{S}$ is the average of all unique source pair distances, $S$ is the set of all sources, and $C$ is the set of all estimated clusters.
 
Table \ref{tab:cts_grouped} shows $\Tilde{d}_{c_x}^{s_z}$ averaged over 200 simulation scenarios for each 4SA and 2SL scenes respectively. A small value of $\Tilde{d}_{c_x}^{s_z}$ for $x=z$ indicates that $\bar{\rho}_{c_x}$ is very close to $\rho_{s_z}$, with $\Tilde{d}_{c_x}^{s_z}=0$ denoting an exact superposition. This goes along with a high value of $\Tilde{d}_{c_x}^{s_z}$ for the other sources where $\Tilde{d}_{c_x}^{s_z} > 1$ indicates that the distance between a source and a cluster centroid is larger than the average distance of sources. For both 4SA and 2SL scenes, we can indeed observe the aforementioned effects for the first four and two columns, respectively. When compared to \cite{Gergen_journal_2015} and \cite{Nelus2021}, which both use unfurnished shoebox rooms, it can be observed that the results generated by the improved version of unsupervised CFL proposed in this work show great potential, especially for the 2SL scene which clearly outperforms \cite{Nelus2021}. 

Moreover, given the distinct characteristics of each combination of speaker groups and ASN constellations, our approach generates a dynamic range of clusters. Since we do not limit the a priori number of clusters/sources and the acoustic environment is reverberant and diverse, CFL identifies clusters of similar weight update vectors not only in the vicinity of sources but also in areas where several sources mix, e.g., hallway ($c_8$) and toilette ($c_7$) clusters in Fig. \ref{fig:apartment_layout}. Thus the clustering results are able to reflect the scenario's acoustic diversity which is a good basis for advanced ASN applications. The number of simulation scenarios $N_{c_x}$ where a cluster $c_x$ is estimated along with its average number of nodes $\overline{N}_{\mathrm{clients},c_x}$ is presented in Table \ref{tab:values_grouped} for both for 2SL and 4SA. This indicates good clustering reliability as at least $N_S$ but no more than $3*N_S$ clusters are frequently generated.

 

Systematic variations of $\lambda$ from (\ref{eq:lambda}) were included in the MV calculation. For the 2SL and 4SA scene the best results were obtained using $\lambda=0.5$ and $\lambda=0$, respectively. This indicates that relying solely on the inter-cluster similarities is sufficient when this uses clusters from more than two sources, otherwise intra-cluster information is also helpful.

 

\begin{table}
	\caption{Number of simulation scenarios generating cluster $c_x$ ($N_{c_x}$) and mean number of clients in $c_x$ ($\overline{N}_{\mathrm{clients},c_x}$). Results include entire SINS apartment (4SA) and living room only (2SL).}
	\centering
	\resizebox{\columnwidth}{!}{%
	\begin{tabular}{l|ccccccccccc}
		\toprule
		                                      4SA & $c_1$   & $c_2$ & $c_3$ & $c_4$ & $c_5$ & $c_6$ & $c_7$ & $c_8$ & $c_9$ & $c_{10}$  & $c_{11}$  \\ \midrule
		$N_{c_x}$                             &  200    &  200  &  192  &  188  &  182  &  163  &  125  & 89   &  52   &  10   &    1           \\
		$\overline{N}_{\mathrm{clients},c_x}$ &  10.4   &  5.1  &  5.2  &  5.1  &  5.2  &  5.7  &  4.8  &  4.6  &  4.1  &   3.1     & 5         \\
		
				\midrule
		                                      2SL & $c_1$ & $c_2$ & $c_3$ & $c_4$ & $c_5$ & $c_6$ \\ \midrule
		$N_{c_x}$                             &  200  &  200  &  79   &  32   &   8 &   4   \\
		$\overline{N}_{\mathrm{clients},c_x}$ &  13.8   &  6.9  &   6.2   &  3.6  &   6.5 &   2.5
	\end{tabular}%
	}
	\label{tab:values_grouped}
\end{table}

\subsection{Gender recognition}
\label{sec:genderrecog}

Gender recognition was proposed as an objective measure for evaluating clustering utility and, indirectly, its performance in both \cite{Gergen_journal_2015} and \cite{Nelus2021}. To allow a direct comparison, we employ the gender recognition model $e$ described in \cite{Nelus2021} which is trained on the Libri-server dataset along with RIR-based data augmentation for 13 epochs using a cross-entropy loss function and an SGD optimizer with $l_r=0.01$. The input feature representation $\gvec{Y}$ is extracted for each 10 s utterance with $L_1=0.064$, $R_1=0.02$, and $K=40$.


The evaluation metrics used are the Accuracy ($A_{cc}$) and $F_1$-score ($F_1$) between the predicted and the ground truth gender labels of a cluster. These are averaged across all clusters in a scenario and averaged again over all 200 simulation scenarios. The predicted and ground truth gender labels of a cluster are given by the mode (majority decision) of the predicted and ground truth gender labels of its nodes. When using MVs, the predicted gender of a cluster is the MV-weighted average of its nodes. The ground truth gender of a node is given by the gender of the source with the shortest first peak delay of the RIR. The predicted gender of a node is the mode of the predicted gender across four utterances (40 s).

To gain a broader perspective on the capabilities and limitations of unsupervised CFL, we propose two meaningful prior-knowledge cases. In the first, we solely rely on the clustering results of the first $N_S$ clusters from Table \ref{tab:cts_grouped}. While this information is unlikely to be available in practice, it provides a good benchmark. In the second case, we select the $N_S$ clusters with the highest gender classification NN output confidence. These might not necessarily be the closest to each source.


 
Results are presented in Table \ref{tab:grouped_gender_acc} for both 4SA and 2SL ASN scenes. As a general observation, it is clear that the proposed MV generation scheme has a great impact on improving the performance of cluster-level gender predictions. Moreover, an MV threshold of $v=0.8$ appears to offer the best overall outcome as it discards some non-helpful nodes. Interestingly, much more aggressive thresholding does not offer better performance as too few nodes remain for gender decision aggregation, once more emphasizing the inherent benefits of clustering. Gender recognition achieves excellent performance for the 2SL ASN scene regardless of the prior knowledge type included. For the 4SA ASN scene, gender recognition is, unsurprisingly, more challenging, and using the $N_S$ closest-to-each-source clusters does offer a small advantage. Nonetheless, it is interesting to see that even when including information from more reverberant clusters, the proposed clustering scheme offers significant advantages to the network-wide gender recognition task. 
\begin{table}
    \caption{Gender recognition Accuracy ($A_{cc}$) and $F_1$-score ($F_1$) of estimated clusters, without and with membership value (MV) weighting using threshold $v$, averaged over 200 scenarios. Closest-to-source (CTS) clusters or clusters with highest confidence (Confidence-based) are selected.}
	\centering
	\resizebox{\columnwidth}{!}{%
		\begin{tabular}{@{}l|lllllll@{}}
			\toprule
			Full apartment (4SA)  & no MV       & $\upsilon = 0$    & $\upsilon = 0.5$ & $\upsilon = 0.8$ & $\upsilon = 0.9$          & $\upsilon = 0.95$  \\ 
			\midrule
			CTS-based $A_{\mathrm{cc}}$ (\%)         & 83.4           & 91.5           & 93.4              & \textbf{93.7}             & 93.2             & 92.7               \\
			CTS-based $F_1$ (\%)                     & 82.4           & 90.1           & 93.1            & \textbf{93.2}             & 92.7             & 92.0               \\
			Confidence-based $A_{\mathrm{cc}}$ (\%) & 83.4           & 92.1           & \textbf{92.8}             & \textbf{92.8}             & 92.2             & 91.4               \\
			Confidence-based $F_1$ (\%)             & 82.4           & 91.3           & \textbf{92.1}             & \textbf{92.1}             & 91.3             & 90.4               \\ \midrule
			
			Living room (2SL)  \\ \midrule               
			CTS-based $A_{\mathrm{cc}}$ (\%)         & 88.1           & 98.5           & 99.0             & \textbf{99.8}             & \textbf{99.8}             & 99.5               \\
			CTS-based $F_1$ (\%)                     & 87.0           & 98.5           & 99.0             & \textbf{99.8}             & \textbf{99.8}             & 99.5               \\
			Confidence-based $A_{\mathrm{cc}}$ (\%) & 88.1           & 98.8           & 99.3             & \textbf{99.8}             & 99.3             & 99.0               \\
			Confidence-based $F_1$ (\%)             & 87.0           & 98.8           & 99.3             & \textbf{99.8}             & 99.3             & 99.0               \\

		\end{tabular}%
	}
	\label{tab:grouped_gender_acc}
\end{table}
\section{Conclusions}
An improved version of unsupervised CFL was proposed in order to handle challenging acoustic environments with multiple simultaneously active sources and minimum training data requirements. The approach is robust and versatile, estimating not only the nearest source-dominated clusters but also those in the reverberant sound field. This, in turn, shows great potential for enhancing additional ASN-deployable tasks such as wake word detection or event classification in future works.

\bibliographystyle{IEEEtran}
\bibliography{refs}

\end{document}